# Superconductivity in Strong Spin Orbital Coupling Compound $Sb_2Se_3$

P. P. Kong[1], F. Sun[1,3], L.Y. Xing[1], J. Zhu[1], S. J. Zhang[1], W. M. Li[1], Q. Q. Liu[1], X. C. Wang[1], S. M. Feng[1], X. H. Yu[1], J. L. Zhu[1,4*], R. C. Yu[1], W. G. Yang[3,5], G. Y. Shen[5], Y. S. Zhao[1,4], R. Ahuja[6], H. K. Mao[3,5], C. Q. Jin[1,2*]

1. Beijing National Laboratory for Condensed Matter Physics and Institute of Physics, Chinese Academy of Sciences, Beijing 100190, China

2. Collaborative Innovation Center of Quantum Matters, Beijing, China

3. Center for High Pressure Science & Technology Advanced Research (HPSTAR), Shanghai, China

4. HiPSEC, Department of Physics and Astronomy, University of Nevada at Las Vegas, Las Vegas, NV 89154-4002, USA

5. High Pressure Synergetic Consortium (HPSynC) & High Pressure Collaborative Access Team (HPCAT), Geophysical Laboratory, Carnegie Institution of Washington, Argonne, Illinois 60439, USA

6. Department of Physics & Astronomy, Uppsala University, Box 516, 75120 Uppsala, Sweden

**Corresponding Authors: Jin@iphy.ac.cn**

**Jinlong@iphy.ac.cn**

**Recently, $A_2B_3$ type strong spin orbital coupling compounds such as $Bi_2Te_3$, $Bi_2Se_3$ and $Sb_2Te_3$ were theoretically predicated to be topological insulators and demonstrated through experimental efforts. The counterpart compound $Sb_2Se_3$ on the other hand was found to be topological trivial, but further theoretical studies indicated that the pressure might induce $Sb_2Se_3$ into a topological nontrivial state. Here, we report on the discovery of superconductivity in $Sb_2Se_3$ single crystal induced via pressure. Our experiments indicated that $Sb_2Se_3$ became superconductive at high pressures above 10 GPa proceeded by a pressure induced insulator to metal like transition at ~3 GPa which should be related to the topological quantum transition. The superconducting transition temperature ($T_C$) increased to around 8.0 K with pressure up to 40 GPa while it keeps ambient structure. High pressure Raman revealed that new modes appeared around 10 GPa and 20 GPa, respectively, which correspond to occurrence of superconductivity and to the change of $T_C$ slop as the function of high pressure in conjunction with the evolutions of structural parameters at high pressures.**

One big breaking through in the studies of strong spin orbital coupling (SOC) system is the discovery of $A_2B_3$ topological insulators (TIs)[1]. TIs is characterized by insulating bulk energy gap like in ordinary insulators but gapless edge or surface states protected by time-reversal symmetry, and therefore trigger wide interests in physical sciences[1-8]. As one of the extraordinary states of topological quantum phases, topological superconductor with Majorana bound state at edge or surface, shows a potential application as well as provides theoretical foundation in topological quantum computations[9]. Superconductivity in TIs was firstly realized in Cu-intercalated $Bi_2Se_3$[10]. In this system, the copper atoms reside in the van der Waals gaps between two $Bi_2Se_3$ layers resulting in the superconductivity with Cooper pairing existing up to 3.8 K. Further efforts on introducing superconductivity into topological insulator and exploring clean topological superconductors without doping were resorted to high pressure[11-13]. Comparing with chemical substitution, applying pressure is an effective tool of inducing novel properties in topological insulators by directly modifying the electronic structure without introducing defects or impurities. For instance, pressure can effectively tune the crystal field splitting and hybridization between Sb and Se[14], resulting the emergence of topological insulator state. The $A_2B_3$-type TIs (A = Sb, Bi; B = Se, Te) except for $Sb_2Se_3$ exhibit rich and novel phenomena under high pressure, such as structure phase transitions[11-13,15-21], insulator-to-metal transitions[13,17], and even superconductivity[11-13,18-22]. The bulk superconducting states of $Bi_2Te_3$ and $Sb_2Te_3$ were observed experimentally as the increase of pressure and were proposed to be likely topologically related[11,12]. Additionally, the high pressure tuned superconductivities were realized in $Bi_2Se_3$[13,22] and $Cu_xBi_2Te_3$[19] system. All these TIs provided possible pathway to search for Majorana fermions through high pressure technique. $Sb_2Se_3$ compound, as a member of $A_2B_3$-type strong SOC semiconductors (A=Sb, Bi; B=S, Se, Te), has attracted great amount of attention in the past few years, due to its switching effects[23], high thermoelectric power[24], and good photovoltaic properties[25]. Unlike $Bi_2Te_3$, $Bi_2Se_3$ and $Sb_2Te_3$, $Sb_2Se_3$ was thought to be topological trivial[1]. However, it was recently theoretically[26,27] predicted that $Sb_2Se_3$ could transform from band insulator to topological insulator under high pressure and thereafter the prediction of topology was experimentally elucidated[14]. Here, we report the effect of pressure on the transporting properties of $Sb_2Se_3$ in combination with structural studies through Raman spectra and *in situ* angle-dispersive synchrotron x-ray diffraction (AD-XRD). Our experiments indicated that $Sb_2Se_3$ transformed from insulator into metal like and further to superconductive state as increase of pressure, during which there was no crystal structure phase transition occurred in our studied pressure range. Two new Raman active modes were observed at high pressure, respectively, where correspondingly the superconductivity emerges and the *$dT_C/dp$* slop changes.

**Results**

**Electronic transport properties of $Sb_2Se_3$ single crystal** $Sb_2Se_3$ is a band insulator with energy gap of $\sim$1 eV[28] and with resistance at the level of $10^9$ Ω at ambient condition. $Sb_2Se_3$ compound undergoes an insulator to metal like transition upon compression, as shown in Fig. 1 (a). As increasing pressure, the resistance decreases over the entire measured temperature range, showing four orders of magnitude drop at room temperature with pressure up to 3.4 GPa, which could be related to the topological quantum transition[27]. The transport must contain conductance contribution from surface but the contribution cannot be differentiated at this moment. From 3.4 to 8.1 GPa, the temperature dependence of resistance is similar to that for topological insulator $Bi_2Se_3$ with an upturn of resistance at low temperature[17]. However, the upturn in resistance is barely visible, due to the high carrier density screening effect[29]. In $Sb_2Se_3$, superconductivity becomes more pronounced with pressure higher than 10 GPa, as shown in Fig. 1 (b). The signature of superconductivity is the resistance drop at around 2.2 K with decreasing temperature under pressure of 10 GPa. Resistance drops more obviously with further increasing pressure and zero-resistance state was fully realized above 15 GPa. Determination of superconducting transition temperature ($T_C$) was based on the differential of resistance over temperature, as elucidated in Ref. [11]. To validate the superconductive character, the evolutions of $T_C$ at 15.4 GPa as a function of magnetic field are performed, as shown in the inset of Fig. 1 (c). The suppression of $T_C$ with increasing magnetic field confirms the superconductivity. Using the Werthdamer-Helfand-Hohenberg formula[30], $H_{C2}(0) = -0.691 \left[ \frac{dH_{C2}(T)}{dT} \right]_{T=T_C} \bullet T_C$, the upper critical field $H_{C2}(0)$ is extrapolated to be 3.9 T with magnetic field $H$ paralleling to $c$ axis of $Sb_2Se_3$ single crystal at 15.4 GPa. Figure 2 (a) shows resistance evolution of $Sb_2Se_3$ compound as the function of pressures at various temperatures. The evolutions of $T_C$ from different experimental runs as functions of pressures are plotted in Fig. 2 (b). $T_C$ increases rapidly to 6 K at 20 GPa with a rate of 0.40 K/GPa. Subsequently, the increase rate of $T_C$ is 0.19 K/GPa and thereafter, $T_C$ increases very slowly to around 8 K at 40 GPa. We further conducted Hall effect measurement with a magnetic field $H$ perpendicular to $a$–$b$ plane of the $Sb_2Se_3$ single crystal and sweep the $H$ at fixed temperatures (2 K, 30 K and 297 K, respectively) for each given pressure. The Hall resistance as a function of applied pressure at 30 K in Fig.1 (d), shows a linear behavior with a positive slope, indicating that $Sb_2Se_3$ is of hole type carrier over the entire pressure range measured. Carrier densities increase almost two orders of magnitude as pressure increases from 10 GPa to 18 GPa (Fig. 2 (c)). For instance, the carrier density changes from ~ $10^{19}$ $cm^{-3}$ to $10^{21}$ $cm^{-3}$ with increasing rate of ~ $6.2 \times 10^{19}$ $cm^{-3}$/GPa from 10 GPa to 18 GPa at both 2 K and 30 K, suggesting significant change in the electronic band structure with pressure[22]. Apparently, the rapid increase of $T_C$ of the superconductive phase from 10 GPa to 18 GPa is intimately related to the increase of carrier density, as observed in $Bi_2Te_3$[21], $Bi_2Se_3$[13,22] or $Sb_2Te_3$[12] systems. The difference is that all these phenomena of $Sb_2Se_3$ occur within the initial ambient crystal structure[31] (see Supplementary Fig. S1 online), other than pressure induced crystal structure changes in $Bi_2Te_3$, $Bi_2Se_3$ or $Sb_2Te_3$ compounds. It should be noticed that further increasing pressure will continuously increase the carrier density in the measured pressure range. The upright dotted lines in Fig. 2 represent the pressures, at which insulator to metal like and further to superconducting transitions occur.

Figure 3 gives the *ac* magnetic susceptibility measurement of $Sb_2Se_3$ single crystal as functions of pressure and temperature. The sudden drop of the real part of *ac* susceptibility appears at pressure above 15 GPa in the measured temperature range (2 K to 300 K), which gets more pronounced with further increasing pressure, strongly indicating the bulk nature of the superconductivity. The $T_C$ increases upon compression, and non-monotonically increased to 39 GPa, which shows consistent trend compared with the $T_C$ obtained from resistance measurement.

**High pressure structural evolution of $Sb_2Se_3$ crystal** To understand the pressure induced novel physical properties in terms of structural mechanism, we conducted *in situ* Raman spectra and angle resolved x-ray diffraction (AD-XRD) measurements at high pressures. The $Sb_2Se_3$ crystallizes in space group *Pbnm*[28,31,32] at ambient condition and sketch of the coordination environment around the Sb1 and Sb2 cations is shown in Supplementary Fig. S2 (a) online. The AD-XRD results (see Supplementary Fig. S1 online) reveal that $Sb_2Se_3$ maintains ambient-pressure phase within the pressure range in our experiment, consistent with the recent observation[31]. At this point, the ambient structure of $Sb_2Se_3$ stabilizes within entire pressure range we studied, indicating that insulator to metal like to superconducting transitions as increase of pressure are type of electronic phase changes related to local structure evolutions. Raman spectroscopy is sensitive to local bond vibrations and symmetric broken[33-36], therefore it can provide evidence for the rich property evolution of $Sb_2Se_3$ under high pressure. As shown in Fig. 4, there are two new vibration modes appeared at 10 and 20 GPa, respectively, which agrees well with the reported results[14,31]. Vibration modes of Raman spectra are fitted by Lorentzian peak configuration with pressure up to ~20 GPa, as plotted in Fig. 4. At ambient pressure, seven modes denoted as *M1* (105 cm$^{-1}$), *M2* (122.6 cm$^{-1}$), *M3* (131 cm$^{-1}$), *M4* (155.2 cm$^{-1}$), *M5* (181 cm$^{-1}$), *M6* (189.6 cm$^{-1}$) and *M7* (213.3 cm$^{-1}$) are clearly seen in Fig. 4 (a). The vibration modes as the function of pressure are fitted to the linear equation (dashed line) $\omega_p = \omega_0 + \frac{d\omega}{dP} P$ (Fig. 5 (d), except for *M5* and *M8*). *M2* and *M5* modes get softened around 2.5 GPa, while *M6* and *M7* modes almost keep constant. At ~10 GPa, *M5* disappears accompanied by a new mode *M8* showing up, where superconductivity occurs. Further increasing pressure to around 20 GPa, another new mode *M9* is observed, which results in the change of increase rate of $T_C$.

**Discussions**

In Fig. 1(b), there appear to be two transitions for the two highest pressures. As in Bi$_2$Se$_3$[22] and Bi$_4$Te$_3$[35] compounds, they also demonstrated two transitions around *BCC* phase coexists with other phase. Considering the non-hydrostatic pressure environment, the high pressure *BCC* phase could be present with the *Pbnm* phase in the resistance measurement at the two highest pressures. However, the pressure gradient in the sample chamber could be the other possible reason of $T_C$ distribution. Figure 5 summarizes the insulator to metal like to superconductivity transitions of Sb$_2$Se$_3$ crystal as a function of pressure through the structure parameters and corresponding electronic property evolution. At ~2.5 GPa, insulator to metal like transition is directly monitored by sharp drop of resistance as shown in Fig. 1 (a). Similar to the three-dimensional time-reversal invariant topological insulators of $Bi_2Se_3$[33] or $Sb_2Te_3$[34] compounds, all of which shows electronic topological transition (ETT) in this low pressure range, our Raman results in Fig. 5 (d) show that the *M2* and *M5* vibration modes gradually soften with pressure up to 2.5 GPa, which agrees well with the $E_g^2$ phonon softening and the asymmetric peak configuration of the linewidth[14]. Specifically, *M2* phonon mode shows a softening and then the sign of slope ($\frac{d\omega}{dP}$) changes and FWHM of *M2* mode is maximum near 2.5 GPa (see Supplementary Fig. S3 online), which may be related to ETT or Lifshitz transition. Our spectra results indicated that during Raman experiments higher laser power will heat and transform the specimen to a new phase as evidenced by vibration mode dramatic change (see Supplementary Fig. S4 online). There is no crystal structural phase transition in the pressure range we studied, and the lattice parameters and volume of our data are consistent with the results in Ref. [31]. However, the anomaly of the bond length in both experiments together with the Raman vibration mode change confirm that the insulator to metal like transition is local structural distortion related. To understand the structural related mechanism of transport properties, we plot the bond angles of ∠Se3-Sb2-Se3 and ∠Se1-Sb2-Se3 to demonstrate their correlation with the insulator to metal like transition and high pressure tuned superconductivity. These two angles increase linearly with pressure up to ~2.5 GPa, then the value of ∠Se1-Sb2-Se3 turns to decrease linearly and the value of ∠Se3-Sb2-Se3 sharply drop to ~91º and increases linearly with pressure to ~10 GPa, as shown in Fig. 5 (b). The *a*/*b* ratio reaches to 1 at 2.5 GPa and followed by parabolic trend with pressure up to 10 GPa. As mentioned above, the crystal field splitting and hybridization must be modulated accordingly. More specifically, the variation of ∠Se1-Sb2-Se3 will mainly rotate the polyhedral of Sb(1)Se$_7$ along the *c* axis while evolution of ∠Se3-Sb2-Se3 will tilt and distort the polyhedral of Sb(1)Se$_7$ at high pressure, as the schematic of the crystal structure shown in Supplementary Fig. S2 (b) and (c) online. At ~10 GPa, the sign of increase rate of ∠Se3-Sb2-Se3 changes, the value of ∠Se1-Sb2-Se3 suddenly increases (as shown in Fig.5 (b)), and a new Raman vibrational mode *M8* at 200 cm$^{-1}$ emerges (as shown in Fig.5 (d)), which should correspond to the metal like state to a superconductive state transition in transport measurement. The values of angles of ∠Se3-Sb2-Se3 and ∠Se1-Sb2-Se3 in Fig. 5 (b) decrease in the pressure range of 10 to 20 GPa with a slope of -0.425 º/GPa and -0.295 º/GPa, respectively, then ∠Se3-Sb2-Se3 change to a smaller decreasing rate and ∠Se1-Sb2-Se3 turns to increase with pressure higher than 20 GPa. At 20 GPa, the change of increase rate of the $T_C$ should be due to anomalies of ∠Se3-Sb2-Se3 and ∠Se1-Sb2-Se3, and the appearance of another new Raman vibrational mode *M9* at ~190 cm$^{-1}$. The dependence of $T_C$ in Sb$_2$Se$_3$ on pressure is similar to that in Bi$_2$Se$_3$, in which pressure-induced unconventional superconducting phase has been reported[13,22]. From 10 GPa to ~20 GPa, the increased carrier density suggests an increased electronic density of states, which promotes an increase of $T_C$. The *a*/*b* ratio stays close to 1 between 10 and 20 GPa followed by a decrease with higher pressure, as shown in Fig. 5 (c). Further increasing pressure to 30 GPa, $T_C$ keeps increasing. Like $Bi_2Se_3$[22], balanced electronic contribution, phonon contribution and some other parameters may lead to the very slow variation from 30 GPa to 40 GPa (see Fig. 2 (b)). However, the relationship between $T_C$ increase (> 20 GPa) and carrier density need to be elaborated and established by further experiments.

The complex structure versus rich properties evolution of $Sb_2Se_3$ as a function of pressure, such as ETT and topological superconducting state, invites further studies from theoretical prediction (electronic structural calculation) to validate the possible topological character.

**Methods**

**Sample synthesis:** $Sb_2Se_3$ single crystal was grown by Bridgeman method. High purity Sb and Se elements are mixed at a molar ratio of 2:3, grounded thoroughly in an agate mortar to ensure homogeneity and then pressed into a cylinder. The mixture was sealed in an evacuated quartz tube under vacuum of $10^{-4}$ Pa and then heated to 850 °C for 24 hours. The quartz tube and sample inside were slowly cooled down to 500 °C with a controlled rate of 10 °C /h, then followed by furnace cooling to room temperature. The crystal quality was assured by x ray diffraction & EDX measurements.

**Measurement of high pressure properties:** The resistance measurement of $Sb_2Se_3$ single crystal under pressure was using the standard four-probe method in a diamond anvil cell (DAC) made of CuBe alloy[11,12,13,18]. The diamond was 300 μm in diameter, center flat with 8 degree bevel out to 500μm. A T301 stainless steel gasket was preindented from a thickness of 250 μm to 40 μm, and a hole was drilled at center with diameter of 150 μm. Fine cubic boron nitride (cBN) powder was used to cover the gasket to keep the electrode leads being insulated from the metallic gasket. The cBN powder was pressed and further drilled into a center chamber with a diameter of 100 μm, in which a $Sb_2Se_3$ single crystal piece in dimension of 60 μm × 50 μm × 30 μm was loaded simultaneously with soft hexagonal boron nitride (hBN) fine powder surrounding it as pressure transmitting medium. Slim gold wires of 18 μm in diameter were used as electrodes. Pressure was determined by ruby fluorescence method[36]. The DAC was placed inside a MagLab system to perform the experiments with an automatic temperature control. A thermometer located near the diamond of the DAC is used for monitoring the sample temperature. The Hall coefficient at high pressure was measured using the van der Pauw method. The *ac* susceptibility was measured using inductance method. Two identical coils were wound with one is placed around diamond anvil and the compensating coil directly adjacent. *Ac* susceptibility measurements were carried out under high pressure at 3.8 Oe rms and 1000 Hz. CuBe gasket was preindented from 500 μm to 250 μm and drilled with a center hole of 350 μm in diameter. The $Sb_2Se_3$ single crystal in dimension of 150 μm × 150 μm × 100 μm was placed into the preindented CuBe gasket chamber with neon loaded as pressure transmitting medium. Ruby spheres placed at the top of sample served as a pressure marker.

**Raman spectroscopy & angle-dispersive powder x-ray diffraction measurement under high pressures;** The monochromatic ADXRD measurements at pressure were performed at the 16BMD beamline of the High Pressure Collaborative Access Team (HPCAT) at Advanced Photon Source (APS) of Argonne National Lab (ANL), using a symmetric DAC. The diamond culet was 300 μm in diameter. The gasket was rhenium which was preindented from 250 μm to 40 μm and drilled with a center hole of 150 μm in diameter. The hole was compressed to 120 μm with neon loaded as a nearly hydrostatic pressure transmitting medium. Ruby spheres were placed around sample to monitor pressure. The x-ray wavelength was 0.4246 Å. The XRD patterns were collected with a MAR 3450 image plate detector. The two-dimensional image plate patterns obtained were converted to one-dimensional $2\theta$ versus intensity data using the Fit2D software package[37]. Refinements of the measured XRD patterns were performed using the GSAS+EXPGUI software packages[38]. The high-pressure Raman experiments were conducted on single crystal $Sb_2Se_3$ using Renishaw inVia Raman microscope with laser wavelength 532 nm and spectral resolution ~1 cm$^{-1}$. The gasket was T301 stainless steel, which was preindented from 250 μm to 40 μm and drilled with a center hole of 120 μm in diameter. Ruby spheres were pressure monitors.

**Acknowledgment**

The work was supported by NSF & MOST of China through research projects. W. G. Y, Y.S.Z, H.K.M acknowledge support by EFree, an Energy Frontier Research Center funded by the U.S. Department of Energy (DOE) under Award DE-SC0001057. HPCAT operations are supported by DOE-NNSA under Award No. DE-NA0001974 and DOE-BES under Award No. DE-FG02-99ER45775, with partial instrumentation funding by NSF. APS is supported by DOE-BES, under Contract No. DE-AC02-06CH11357.

**Authors contributions**: CQJ conceived & coordinated the work; XLY grown single crystals with preliminary characterizations; PPK, JZ, SJZ, WML, QQL, XCW, SMF, RCY conducted the high pressure transport measurements; FS, XHY, WGY contributed to the measurements of high pressure structures with the helps of YSZ, GYS; RA contributed to the theoretical analysis with help of JLZ; CQJ, PPK JLZ analyzed the data; CQJ, PPK, JLZ, wrote the paper. All authors contributed to the discussions of the work.

Additional Information: The authors declare no competing financial interests.

**Figure Captions:**

**Figure 1** Electronic transport properties of $Sb_2Se_3$ single crystal. (a) Temperature dependence of *a-b* plane resistance below 8.1 GPa. (b) The *a-b* plane resistance as a function of temperature at various pressures showing a superconducting transition around 2.2 K at 10 GPa. (c) The superconducting transition of $Sb_2Se_3$ with applied magnetic field $H$ perpendicular to the *a-b* plane of the $Sb_2Se_3$ single crystal at 15.4 GPa. The inset shows $T_C$ evolution as a function of magnetic field $H$. (d) The Hall resistance of $Sb_2Se_3$ as a function of applied pressure at 30 K, showing a linear behavior with a positive slope.

**Figure 2** (a) Resistance evolution of $Sb_2Se_3$ single crystal as a function of pressure at fixed temperatures (280 K, 200 K, 2 K), in which insulator, metal like and superconductor are denoted. (b) Pressure dependence of superconductive transition temperature $T_C$ of $Sb_2Se_3$. Solid lines are guides to the eye. Increasing rate of $T_C$ decreases upon compression. (c) Pressure induced changes of carrier density at temperature of 2 K, 30 K and 297 K, respectively.

**Figure 3** The real part of the *ac* susceptibility of $Sb_2Se_3$ single crystal as a function of temperature at various pressures. The inset is the $T_C$ as a function of pressure.

**Figure 4** Pressure evolution of Raman spectra ($\lambda$ = 532 nm, T = 300 K). From low to high wavenumber, phonon modes are denoted as *M1*, *M2*, *M3*, *M4*, *M5*, *M6* and *M7* at 0.9 GPa. The black squares are experimental data points, the red solid lines are the total fitting curves from the sum of the individual Lorentzian fits (blue peaks) to the experimental data points. Arrows mark the appearances of new vibration modes.

**Figure 5** The relation of electrical properties and structure as a function of pressure. (a) $T_C$ from two individual resistance measurement experiments, (b) the angles of ∠Se1-Sb2-Se3 and ∠Se3-Sb2-Se3, (c) *a/b* ratio and (d) Evolutions of Raman vibration modes as functions of pressures. The vertical shadowed boxes indicate the pressure for insulator-metal like-superconductors transition and the change of $T_C$ slop as the function of high pressure, and all the solid and dotted lines are guidelines for your eyes.

**Figure 1**

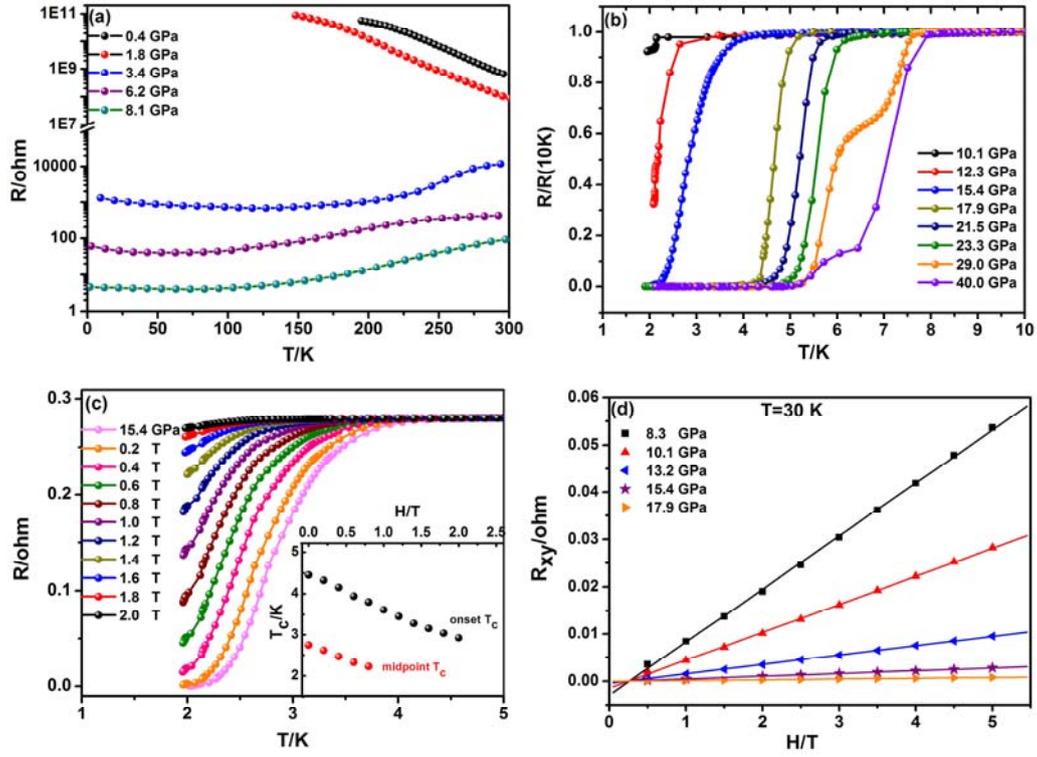

**Figure 2**

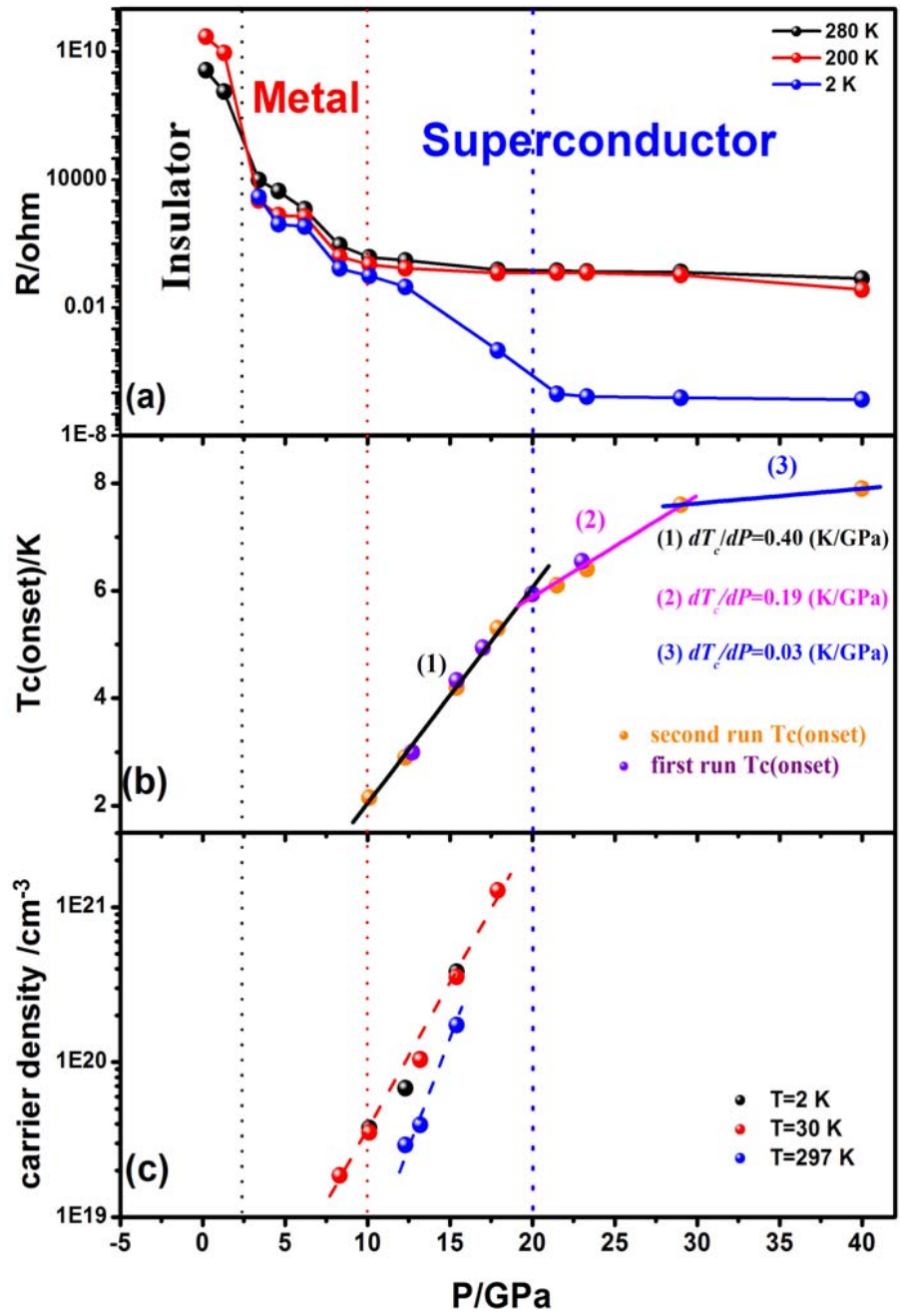

**Figure 3**

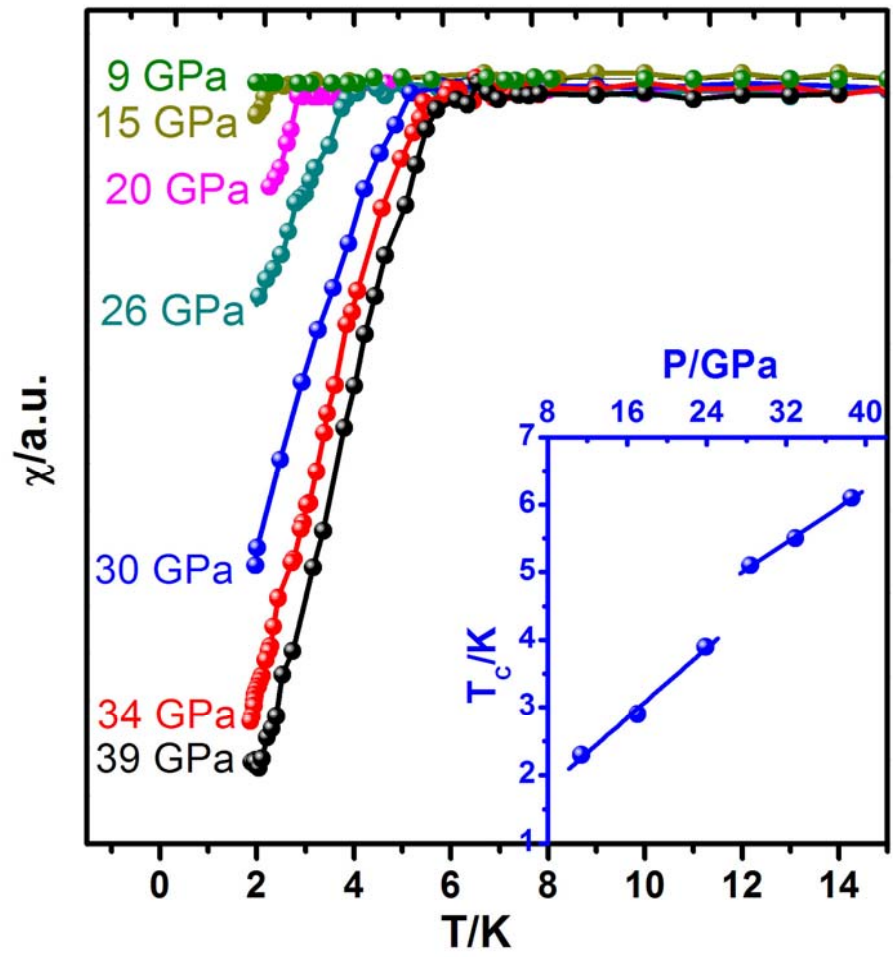

**Figure 4**

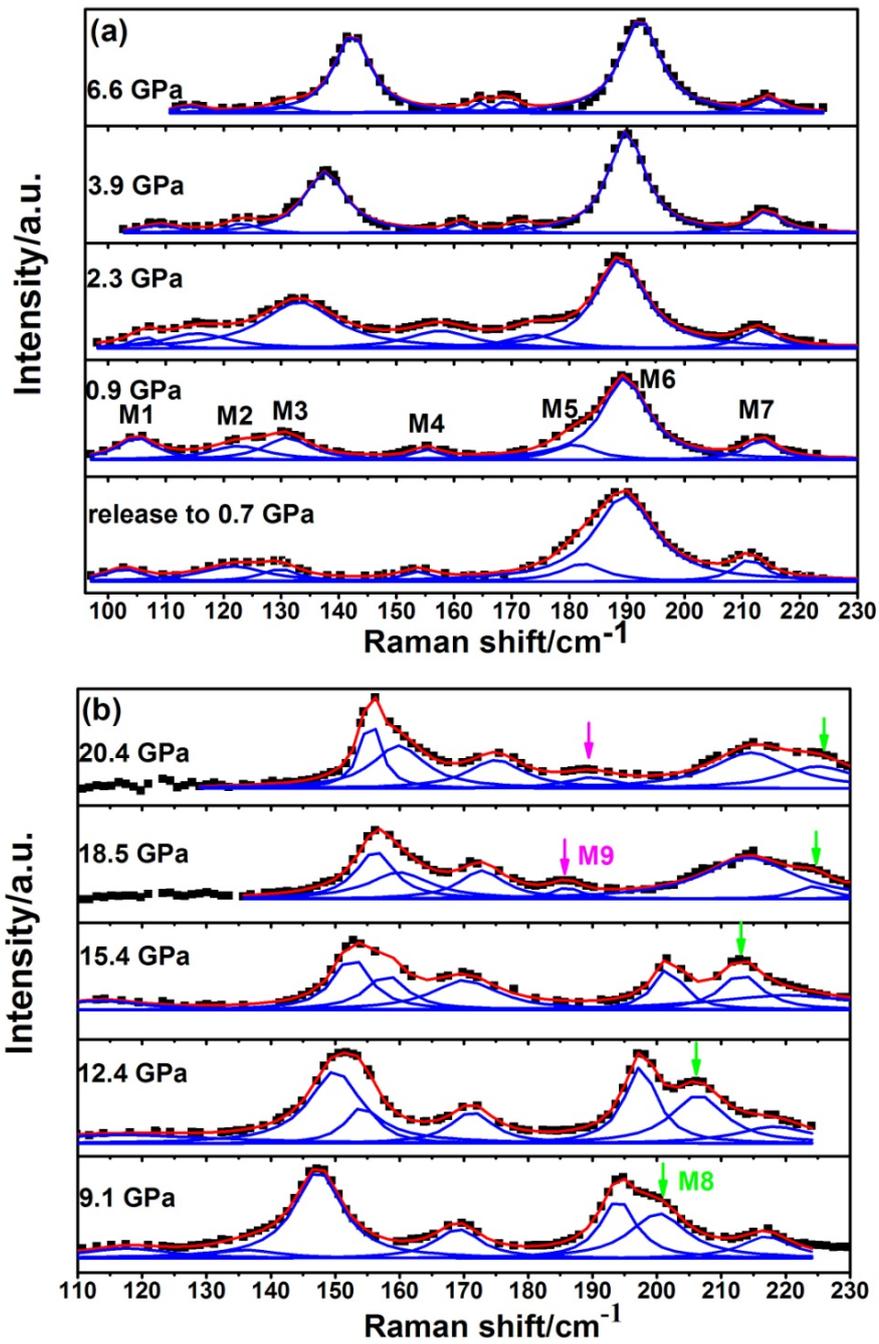

**Figure 5**

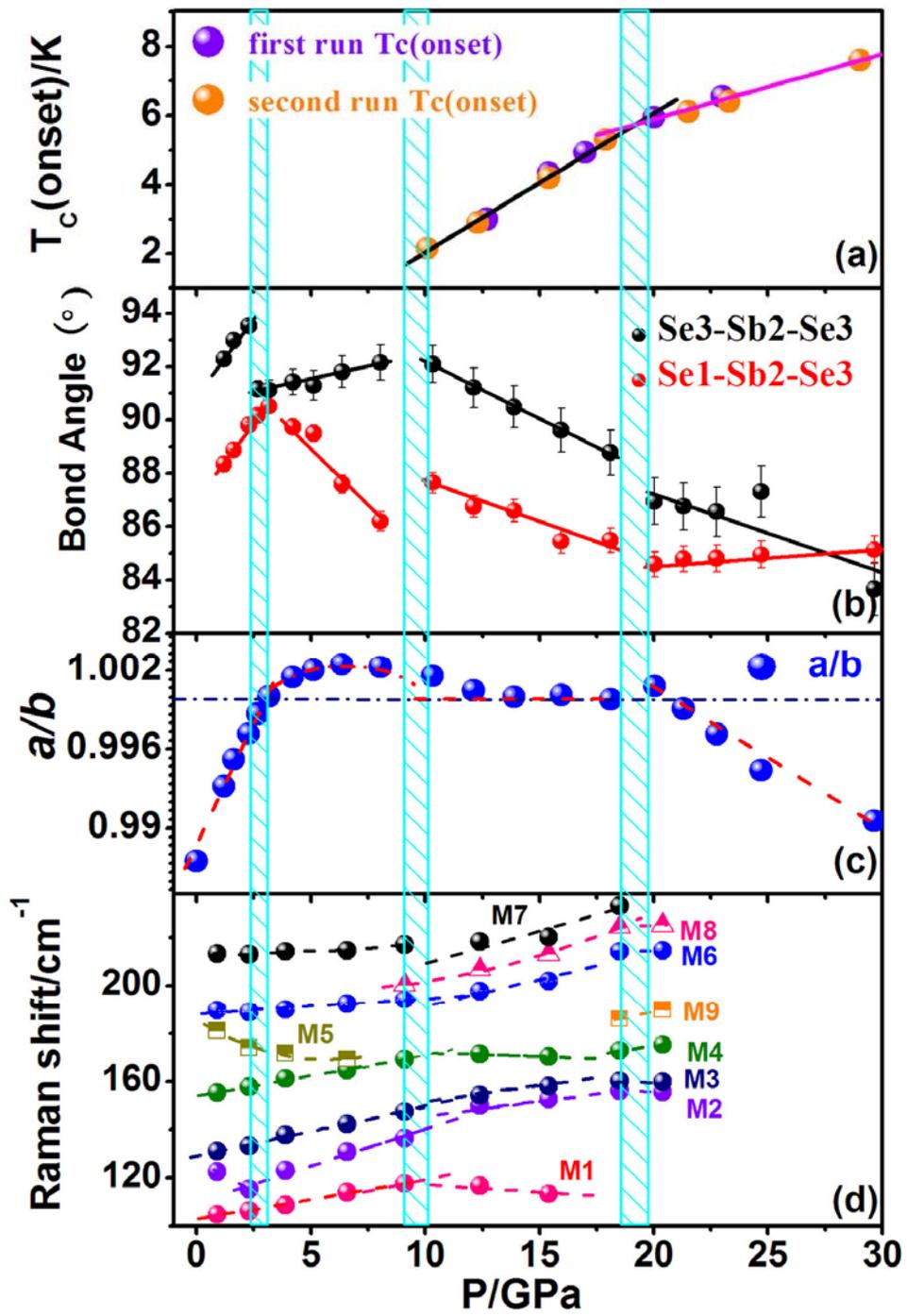

# Superconductivity in Strong Spin Orbital Coupling Compound Sb$_2$Se$_3$

P. P. Kong[1], F. Sun[1,3], L.Y. Xing[1], J. Zhu[1], S. J. Zhang[1], W. M. Li[1], Q. Q. Liu[1], X. C. Wang[1], S. M. Feng[1], X. H. Yu[1], J. L. Zhu[1,4*], R. C. Yu[1], W. G. Yang[3,5], G. Y. Shen[5], Y. S. Zhao[1,4], R. Ahuja[6], H. K. Mao[3,5], C. Q. Jin[1,2*]

1. Beijing National Laboratory for Condensed Matter Physics and Institute of Physics, Chinese Academy of Sciences, Beijing 100190, China

2. Collaborative Innovation Center of Quantum Matters, Beijing, China

3. Center for High Pressure Science & Technology Advanced Research (HPSTAR), Shanghai, China

4. HiPSEC, Department of Physics and Astronomy, University of Nevada at Las Vegas, Las Vegas, NV 89154-4002, USA

5. High Pressure Synergetic Consortium (HPSynC) & High Pressure Collaborative Access Team (HPCAT), Geophysical Laboratory, Carnegie Institution of Washington, Argonne, Illinois 60439, USA

6. Department of Physics & Astronomy, Uppsala University, Box 516, 75120 Uppsala, Sweden

**Corresponding Authors: Jin@iphy.ac.cn**

**jlzhu04@iphy.ac.cn**

**Figure S1 crystal parameter evolutions of Sb$_2$Se$_3$ crystal under pressure** (a) AD-XRD patterns of Sb$_2$Se$_3$ at room temperature with pressure up to 30 GPa. The wavelength is set at 0.4246 Å. (b) Rietveld refined XRD pattern of Sb$_2$Se$_3$ at 10.3 GPa. Hollow dots correspond to the measured spectrum and the red solid line represents the best fitting. The difference spectrum between the measured and the refined pattern is depicted by blue line. Red vertical ticks mark the Bragg peak positions. (c) Pressure dependence of lattice parameters a, b and c, as well as volume V. Birch–Murnaghan equation of state fitting indicates that the bulk module is 32.7 (8) GPa with B$_0$' = 5.6 (4), which is consistent with the value in Ref. [1].

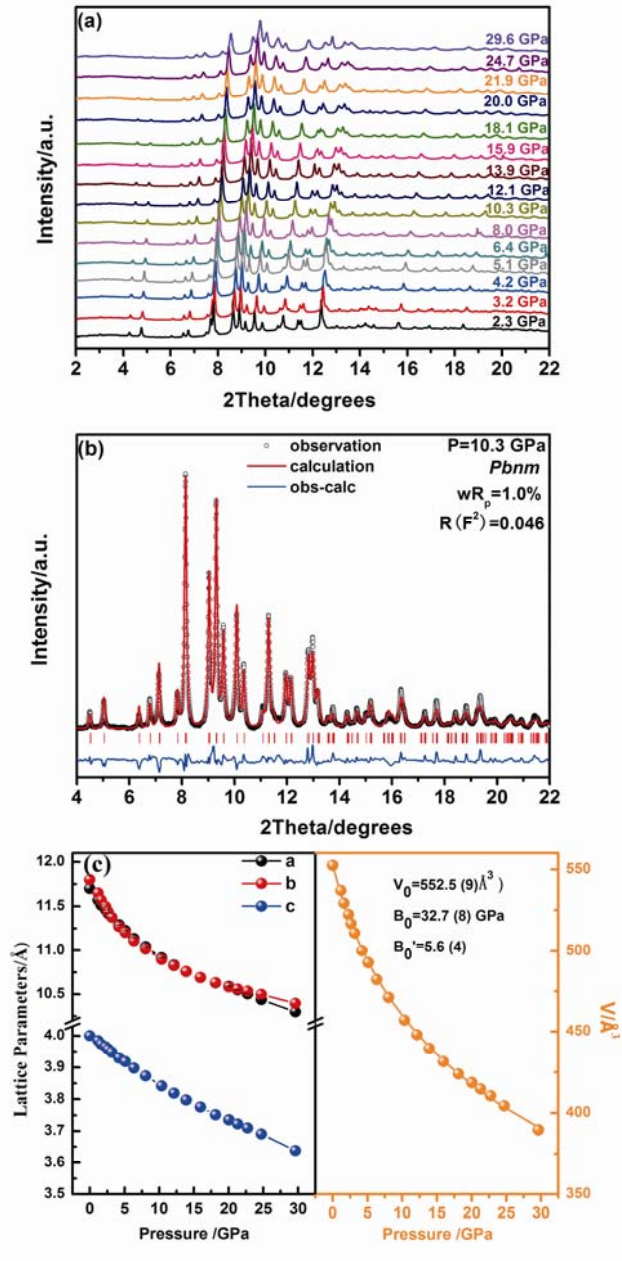

**Figure S2 Schematic show of crystal structure** (a) Sketch of the coordination environment around the Sb1 and Sb2 cations. The Sb1 and Sb2 cations are coordinated by 7 and 7+1 Se anions[1], respectively. Sb1, Sb2, Se1, Se2 and Se3 ions are denoted as magenta, blue, cyan, green and orange spheres, respectively. ∠Se1-Sb2-Se3 and ∠Se3-Sb2-Se3 are denoted by blue and orange arcs, respectively. (b) Top view of *a-b* plane suggests that the decrease of ∠Se1-Sb2-Se3 (< 90°) will mainly rotate the polyhedral of $Sb(1)Se_7$ (magenta polyhedra) normal to the c-axis. Green arrows denote moving direction of Se3 atoms while ∠Se1-Sb2-Se3 decreasing. (c) Schematic show along other direction proposes that the decreases of ∠Se3-Sb2-Se3 will tilt and distort the polyhedral of $Sb(1)Se_7$ (magenta polyhedra). Yellow arrows denote moving direction of Se3 atoms with decreasing ∠Se3-Sb2-Se3.

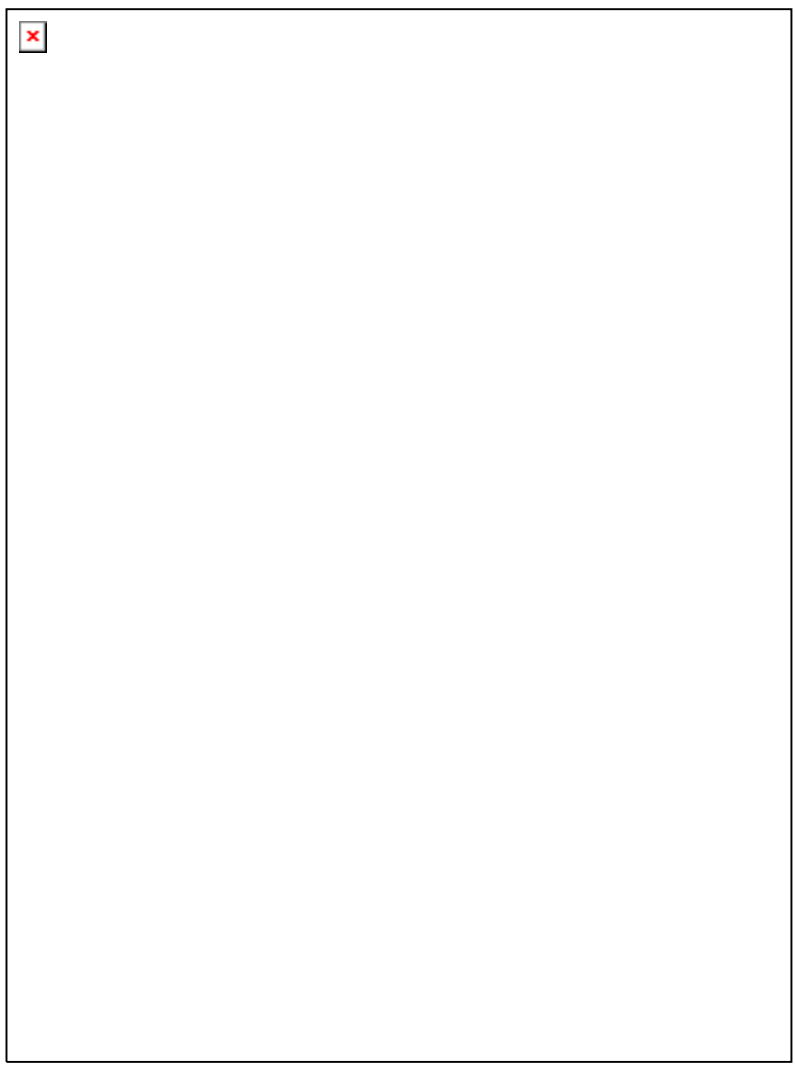

**Figure S3 Configuration of peak width of *M2* Raman mode** Full width at half maximum (FWHM) of the *M2* mode (solid squares) as a function of pressure and the solid line is drawn as guide to the eyes. FWHM of the *M2* mode has two anomalies located at around 2.5 GPa and 10 GPa, respectively.

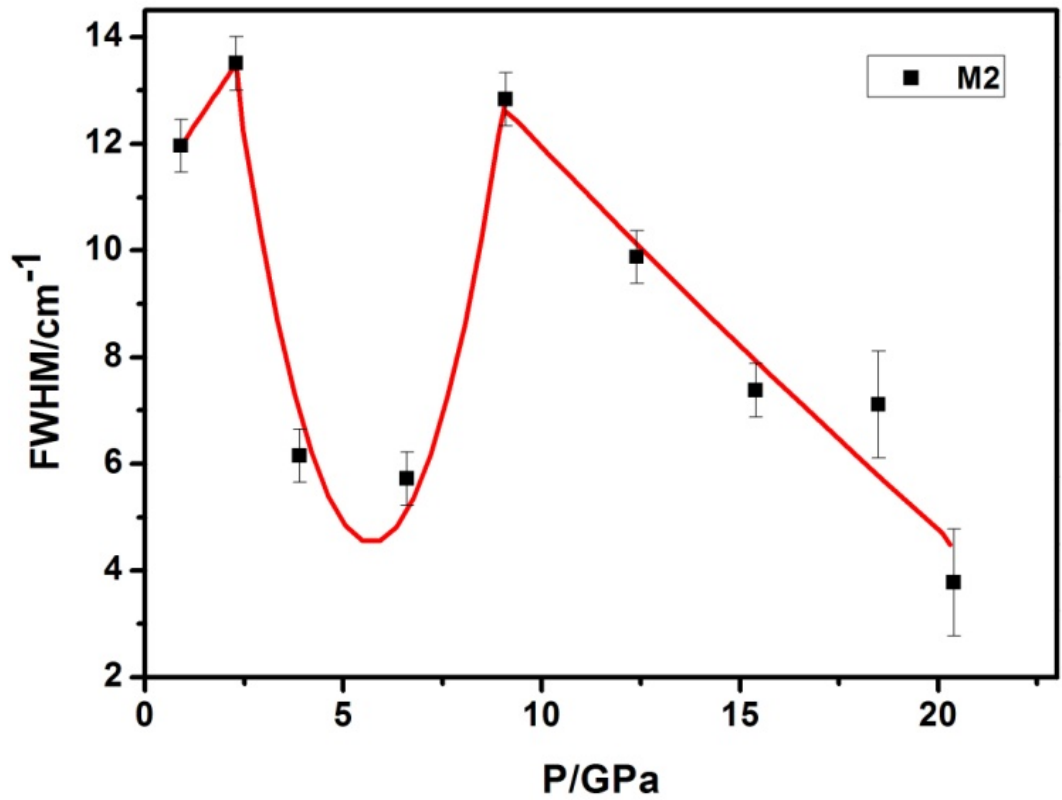

**Figure S4 Heating effect of Raman laser power at atmospheric conditions** The Raman spectrum with high laser power is analogous to Efthimiopoulos's report[1]. The Raman spectrum with low laser power is similar to Bera's report[2]. The inset is the Lorentzian fitting to the Raman spectrum collected at low laser power, the phonon modes are denoted the same as that in high pressure Raman spectra. Seven phonon modes are *M1* (100.3 cm$^{-1}$), *M2* (118.6 cm$^{-1}$), *M3* (129.7 cm$^{-1}$), *M4* (154.6 cm$^{-1}$), *M5* (184.3 cm$^{-1}$), *M6* (191.3 cm$^{-1}$), *M7* (213.1 cm$^{-1}$), respectively.

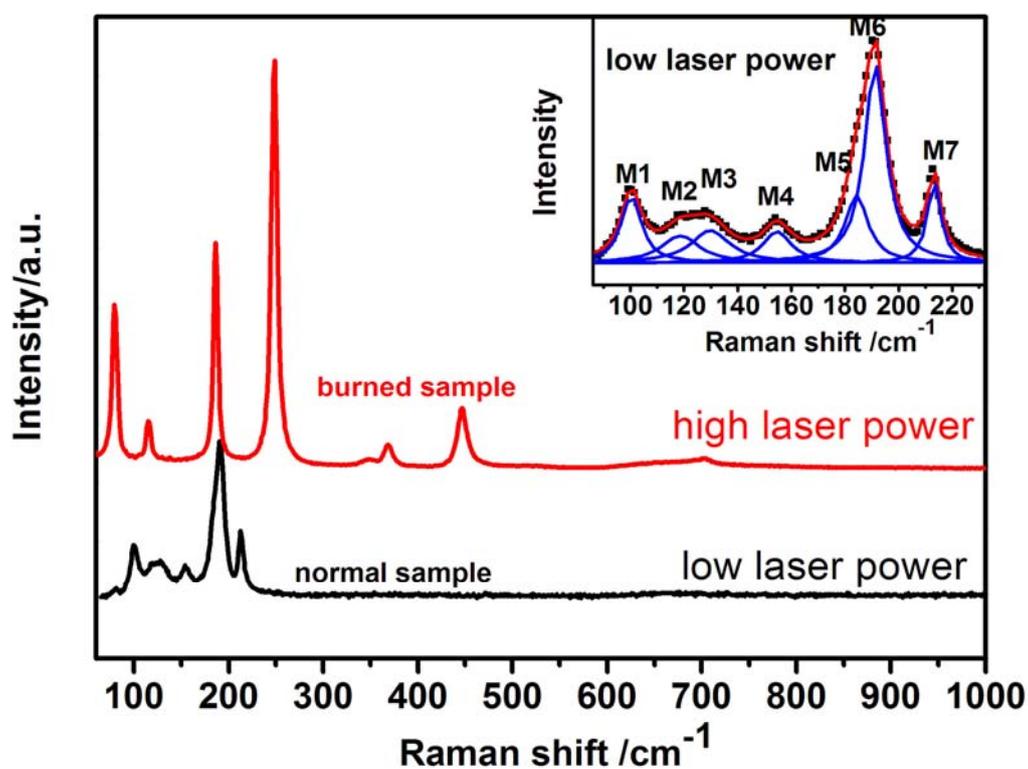